\documentclass[12pt]{article}
 \usepackage{fullpage}
\usepackage{graphicx}
\usepackage{caption}
\usepackage{rotating}
\usepackage{subcaption}
\usepackage{multirow}

\title{A Decentralized and Autonomous Model to Administer University Examinations}
\author{Yogesh N Patil\footnote{The author is working as a Director, Board of Examinations and Evaluations.}, Arvind W Kiwelekar, Laxman D Netak, 
Shankar B Deosarkar \footnote{The author has worked as a Controller of Examinations for 15 years.}\\
Department of Computer Engineering  \\
Dr. Babasaheb Ambedkar Technological University\\ Lonere-402103 India\\
\{ynpatil, awk, ldnetak, sbdeosarkar\}@dbatu.ac.in}

\begin{document}

This is a pre-print of the following Chapter:   Patil Y.N., Kiwelekar A.W., Netak L.D., Deosarkar S.B. (2021) {\bf A Decentralized and Autonomous Model to Administer University Examinations.} In: Lee SW., Singh I., Mohammadian M. (eds) Blockchain Technology for IoT Applications. Blockchain Technologies. Springer

{\bf Cite this chapter as:}
Patil Y.N., Kiwelekar A.W., Netak L.D., Deosarkar S.B. (2021) A Decentralized and Autonomous Model to Administer University Examinations. In: Lee SW., Singh I., Mohammadian M. (eds) Blockchain Technology for IoT Applications. Blockchain Technologies. Springer, Singapore. 

\newpage

\maketitle
\abstract{
	Administering standardized examinations is a challenging task, especially for those universities for which colleges affiliated to it are geographically distributed over a wide area. Some of the challenges include maintaining integrity and confidentiality of examination records, preventing mal-practices, issuing unique identification numbers to a large student population and managing assets required for the smooth conduct of examinations.   These challenges aggravate when colleges affiliated to universities demand academic and administrative autonomy by demonstrating best practices consistently over a long period.
	
In this chapter, we describe   a  model for decentralized and  autonomous examination system
to provide the necessary administrative support. The model is based on two emerging technologies of Blockchain Technology and Internet of Things (IoT). We adopt a software architecture approach to describe the model.  The prescriptive architecture consists of  {\em architectural mappings} which map functional and non-functional requirements to architectural elements of blockchain technology and IoT.    In architectural mappings,  first, we identify common use-cases in administering standardized examinations. Then we map these use-cases to the core elements of blockchain, i.e. distributed ledgers, cryptography, consensus protocols and smart-contracts and IoT. Such kind of prescriptive architecture guide downstream software engineering processes of implementation and testing.}
	 
\section{Introduction}

Information and Communication Technologies (ICT)  have been revolutionizing education systems around the world. Adoption of online mode of teaching and learning by deploying Learning Management Systems \cite{mcgill2009task} and Massive Online Open Courses\cite{breslow2013studying} are some of the indicators of the impact of ICT on education. Web platforms and mobile devices are the technologies which universities are preferring to create and distribute educational contents. 

In recent times, emerging technologies such as Blockchain and Internet of Things (IoT) have opened up opportunities for educators to devise technological interventions to their pressing problems. Some of the novel applications of blockchain technologies include certificates management, evaluating students’ professional ability, and enhancing students’ interactions in e-learning. The detailed surveys of applications of Blockchain technology  \cite{alammary2019blockchain} and IoT \cite{al2020survey} in  Education highlight the importance of these technologies to improve educational outcomes. 

In this chapter,  we present a high-level design of a model to provide administrative support for conducting university examinations around these two emerging technologies of Blockchain and IoT.
Section 2 and 3 describe the university examinations system practised in  India and challenges faced to conduct examinations. In Section 4 and 5, we identify functional and non-functional requirements, especially security and monitoring requirements, to develop Blockchain and IoT based governance support. We adopt a software architecture approach to define the proposed high-level design.   We describe in Section 7 a mapping of architecturally significant requirements to the main design elements from Blockchain and IoT.

\section{University Examinations System in India} Institutes of Higher Education in India can be classified into three broad categories depending upon the level of autonomy granted to them concerning the conduct of examinations. These are: (i) Affiliating Universities, (ii) Autonomous Colleges, and (iii) Deemed Universities.
\begin{enumerate}
    \item {\bf Affiliating University} These are the institutes of higher education to which other colleges are affiliated. Further, affiliating universities prescribe a curriculum, conduct examinations, maintain academic records of students, and award degree certificates to them. The colleges are typically distributed over a wide geographical area sometimes spanning over thousands of kilometres. All the administrative and academic powers are centralized at the university level with little or no autonomy to the colleges affiliated to the university.
    
    These kinds of universities adopt {\em standardized examinations} as a mode of assessment. In standardized exams, thousands of students are assessed through a standard question paper. Unlike the online method of tests consisting of Multiple-Choice-Questions (MCQ),  these examinations consist of essay type descriptive questions, and they are typically conducted in paper-pen mode.

    \item {\bf Autonomous Colleges} Autonomous colleges have autonomy concerning prescribing curricula and conducting examinations. Like affiliated colleges, autonomous institutes are affiliated to a university. For these institutes, the role of a university is to maintain academic records of students and award degree certificates to them. Such autonomous colleges have no powers to award degree certificates.
    
    \item {\bf Deemed Universities} A deemed university is an institute of higher education with power to prescribe curricula, conduct examinations, and award degree certificates. But no control of affiliating colleges to it beyond its campus. 
    
\end{enumerate}
 In this chapter, we propose a governing model to administer standardized examinations for affiliating universities to which autonomous and non-autonomous colleges are affiliated. The proposed model is designed around two technologies of Blockchain and IoT.
 
 \section{Challenges in Administering University Examinations} \label{uc}
 
 Universities face various kind of challenges concerning the conduct of examinations. These  challenges are  of four types, namely, (i) Administrative, (ii) Infrastructure, (iii) Resource, and  (iv) Security \cite{aggarwalCommitteeReport}.   We describe some of the security challenges needing technological interventions in this section. 
 \begin{enumerate}
     \item {\bf Leakage of Sensitive Information [SC1]} Many universities rely on manual processes to transfer sensitive documents such as question papers, identities of paper setters, mark-statements, and stationery items related to the conduct of examinations. Leakage of such sensitive documents before the scheduled examinations by agencies involved in this process have been reported at numerous times. Such leakage has serious consequences on the scheduling of examinations, cancellation of examinations, and repeating examinations. Honest and sincere students need to bear the burden of such leakages.
     \item {\bf Tampering with Valuable Information [SC2]} Incidences of tampering of valuable information such as students grades and the names of students on grade-sheets by administrative staffs engaged in the process of examinations have also been reported. Such incidences lead misinformation and wrong academic credentials to students.
     \item {\bf Generating and Distributing Fake Certificates [SC3]} Generating fake degree certificates for the students who have not enrolled in any degree programs by a third party is also a common threat faced by many universities.  Such incidences are increasingly happening for programs with high job potentials.
     \item {\bf Security threats due to the deployment of ICT [SC4]} Some of the universities have started adopting ICT for effective administration purposes. In addition to the existing security challenges involved in the manual process, the use of ICT has exposed universities to various kinds of cyber-attacks, such as a denial of service attacks. 
 \end{enumerate}
 Addressing these challenges is one of the main motivations behind exploring the applications of  blockchain technology to improve the functioning of examinations system. 
 
 \section{Functional Requirements for University Examinations} \label{fr}
 An examinations section of a university conducts various kinds of examinations throughout a year. These examinations are either mid-semester, end-semester or remedial examinations. The broad process of examinations can be broken down among the following tasks.  To use the Software Engineering terminology, these are the functional requirements or use-cases to be supported by the proposed  system.
\begin{enumerate}
    \item  {\bf To enroll  students for examinations [FRQ1]} Students perform this activity to enroll and express their willingness to appear for a particular course examination. Students earn specified course credits by qualifying a course examination. 
     \item  {\bf To assign identification credentials to students [FRQ2]}.The university assigns a unique identification when a student enrolls for the first course. The unique identification needs to hide students real-life identity at the same time it needs to be linked to the real-life identity.
     \item {\bf To make classroom attendance mandatory for university examinations [FRQ3]:} Many universities make it compulsory to fulfil the requirement of minimum classroom attendance for appearing to the End-Semester-Examinations. Universities need a mechanism to implement this requirement  and to link a students' classroom attendance to the issue of their hall tickets or admit cards [FRQ4]. 
    \item {\bf To issue hall tickets or admit cards [FRQ4]} Hall tickets or admit cards are the paper-based permission documents which are required to enter in an examination hall. It contains photo-identity of a student and courses enrolled by the student.
    
    \item  {\bf To generate  question papers for examinations [FRQ5]} Test examinations for a course can be generated automatically, or course expert can manually prepare it. An automatic system makes use of pre-defined question item bank and set of rules to produce examinations as discussed in \cite{kale2013algorithm,wankhede2016qualitative}. In a manual mode, a course expert submits a question paper to a university.
    \item {\bf To distribute Examinations related material [FRQ6]:} Universities distribute pre-printed stationery to affiliated colleges along with question papers so that affiliated colleges can conduct examinations on-time without any glitches.
    \item {\bf To maintain examination record [FRQ7]:} Universities need to manage the data about the academic credentials achieved by a student during the registration time and after graduation for a long time for verification and validation purposes.
    \item {\bf To issue degree certificates and transcripts [FRQ8]:} Universities award degree certificates to students upon successful completion of a program requirement. Universities issue transcripts are showing details of academic credentials to students.
    \item {\bf To verify and validate students records [FRQ9]:}  In case of students graduated, universities receive requests from various recruitment agencies to validate degree certificates and transcripts. 
    
    \item {\bf To manage assets and inventory of stationary items required for conducting university examinations: [FRQ10]} Universities maintain an inventory of stationery items (e.g., papers, pockets, staplers), and assets like printers, copiers, and vehicles.   Continuous monitoring of the usages of these assets and inventory items is an essential and time-consuming activity to conduct examinations.
   \end{enumerate}

A software system realizing these functional requirements has to share and distribute the sensitive information among various stakeholders  over the internet. Hence, it  necessitates to analyze the design of software system from information security point of view.

 \section{Information Security Requirement Analysis for University Examinations} \label{nfr}
 Many universities have started adopting ICT to provide administrative support for the examinations tasks listed in Section \ref{uc}. These approaches  either outsource the implementation of these functionality to a cloud service provider or realize it as a  centralized client-server application. Though both the approaches bring a degree of automation and efficiency in performing tasks, they are prone to information security attacks. So far no such incidences have been reported but occurrences of such attacks can not be ruled out in future.
 \begin{enumerate}
     \item {\bf Data Confidentiality [SRQ1]:} This requirement ensures that sensitive information such as question papers, model answer, student assessment records, information about affiliated colleges shall not be shared with any unauthorized applications or users. This requirement protects systems against attacks such as intentional and unintentional information leakage and data breaches.
     
     \item {\bf Data Integrity [SRQ2]:} Data integrity is another important requirement to avoid tampering with data. An un-authorized application shall not modify or tamper with information records such as grades of students, enrolled courses and other such information.
     \item{\bf Authentication [SRQ3]:} Authentication of users and applications is the most primitive way to recognize legal users and applications. A strong application protects the system from data breaches and information leakage.
     \item{\bf Authorization [SRQ4]:} Illegal tampering and modification of data shall be  ensured by providing various levels of authorization corresponding to the kinds of users of systems such as teachers, paper setters, evaluators, heads of the departments, principals,and a controller of examinations.
     \item {\bf Denial of Service Attacks [SRQ5]:} Such kinds of attacks are initiated by external users through flooding the network with excessive traffic. For example, when university declares the results of an examination un-intentional denial-of-service attacks can occur if  a large number of students open multiple sessions to know results. A malicious users can intentionally block the access to an examination service by flooding network with excessive traffic.
     \item {\bf Man-in-the-middle [SRQ6]:} In Man-in-the-middle kinds attacks,  a malicious user intercepts communication between two parties with an intention to tamper the messages. From the examination system point of view, a proxy student may appear for an online examination, or modify a question paper delivered over the internet through such kinds of attacks. A robust examinations system shall prevent such kinds of attacks.
     \item {\bf Privilege escalation [SRQ7]:}A malicious user often an authenticated user can use flaws in the system design to gain privileged access to perform unauthorized actions. For example, an authenticated student acquires rights of teacher to modify grade-sheets.
     \item {\bf Privacy Leakage [SRQ8]:}  These are data breach attacks initiated normally by internal staff having access to private information of users. In such kinds of attacks, internal staff shares the private information to third party. In the context of examination system, social profile of students, asset information of the colleges affiliated to the universities are some of the private information. 
 \end{enumerate}
 
 A robust and trustworthy platform is required to implement such kinds of security requirements.
 
 \section{Architectural Elements in  Blockchain} \label{ae}

The existing web technologies have introduced an information-centric model which has revolutionized the education system. For example, the emergence of MOOC \cite{mcauley2010mooc} platforms such as Coursera ($www.coursera.org$) has been attributed to the growth and widespread adoption of online technologies for teaching and learning processes. This centralized model has bridged the information gap that exists between a content provider and content consumers by creating a third-party for information exchange called intermediaries or agents or service providers. 

Some of the advantages of web technologies for content distribution are that processes of teaching and learning are simplified, and the time required to learn has been drastically reduced because of access to quality learning material \cite{kiwelekar2020architectural}.  Despite the various benefits of web technologies, the Internet has always remained an unreliable platform to share valuable personal information because of its mediator-centric model for information exchange.   The information shared with such sites is still susceptible to breach of security and privacy attacks. Due to these limitations,  applications of web technologies have not been extended to other relevant academic processes such as examinations.

The {\em emerging blockchain technology} removes these pitfalls by laying a trust layer on top of the existing Internet technology. It replaces the mediator-centric model of information exchange with the peer-to-peer model or decentralized model.   It transforms the Internet into a trustworthy platform for doing business when transacting parties that do not trust each other. 
Initially emerged as a platform to exchange digital currency over the Internet, now the blockchain technology is gradually emerging as a general-purpose platform for sharing and protecting information \cite{crosby2016blockchain}.  

This section reviews the four fundamental concepts common across various blockchain implementations. These are: \cite{dinh2018}:
(i) Distributed Ledger [DL],
(ii) Cryptography (CRYPT),
(iii) Consensus Protocols (CP), and
(iv) Smart Contracts (SC)

\subsection{Distributed Ledgers} 

In the context of a blockchain-based information system, ledgers are the databases storing up-to-date information about business transactions. These are distributed among all the nodes participating in the network.  So multiple copies of a ledger exist in a business network.  A consensus protocol ensures consistency of each copy with each other. These ledgers are used to store information about valuable assets such as coins, land records, diamonds, student's academic credentials and others.

\begin{figure}[t]
\centering
    \includegraphics[scale=0.28]{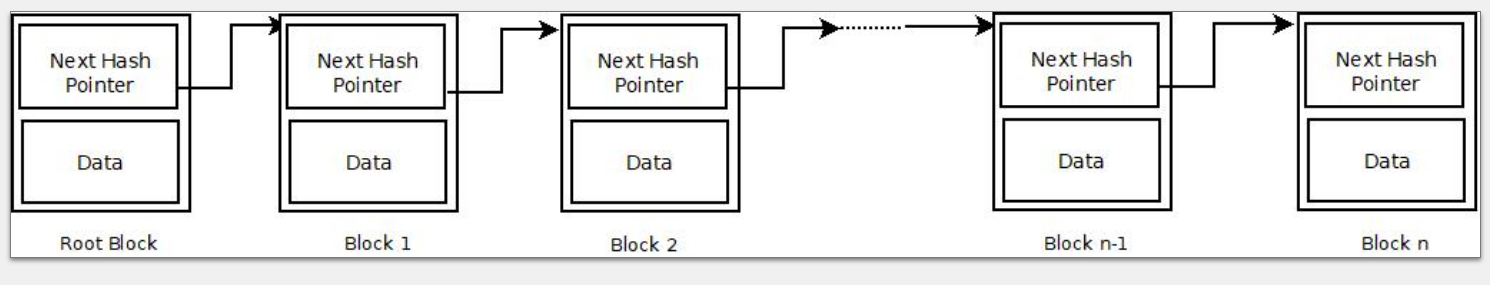}
    \caption{Blockchain}
    \label{bc}
\end{figure}

As shown in Figure \ref{bc}, distributed ledgers arrange records in a chain-format.    Here, multiple transactions related to an asset are grouped in a block. The $(n+1)^{th}$ block in the chain links to the $n^{th}$ block and the $n^{th}$ block links to the $(n-1)^{th}$ block and so on.  Due to this peculiar arrangement, distributed ledgers are also known as a blockchain.  The blockchain data structure permits only append of new records. Updating and deletion of records are not permissible.

The most critical design feature of blockchain-based information system is the use of hash pointers instead of physical memory based pointers to link blocks in a chain. A hash pointer is a message digest calculated from the information content of a block. Whenever a node attempts to tamper the information content, a small change in the information leads to a ripple effect of changes in hash-pointers.  Thus, making it impossible to change the data once it has been recorded in the blockchain. 

The distribution of ledgers across the network and use of hash-pointers enable mediator-less business transactions and immutability of stored information. These are the two significant quality attributes associated with blockchain-based information systems. 

\subsection{Cryptography}\label{crypt}
Blockchain technology makes heavy use of cryptographic functions to assure trust among the users transacting over a blockchain-based business network. These cryptographic functions address various purposes. Some of them are:

1) {\em Authenticating the identity of agents involved in a business transaction}:\newline Blockchain-based systems use a kind of asymmetric key cryptography. These protocols use two different keys called public and private keys. The public keys are open and used as addresses for performing business transactions while private keys are secret and used for validating the transactions. Secure Hash Algorithm (SHA-256 used in  Bitcoin \cite{nakamoto2019bitcoin}) and Elliptic Curve Digital Signature Algorithm   ( ECDSA used in  Hyperledger \cite{cachin2016architecture}) are some of the cryptographic protocols used for this purpose \cite{luntovskyy2018cryptographic}.  Cryptographic functions such as digital signature are also used to authenticate a particular transaction.

2) {\em Ensuring Privacy}: The blockchain technology adopts various mechanisms to preserve the privacy of a transaction.  Below we discuss these mechanisms and their intentions behind the design.  
\begin{enumerate}
    \item  {\em Decentralised Privacy}.   The blockchain technology adopts decentralization as one of the guiding design principles. It eliminates the role of mediator to store transaction information at a central place.  The transaction information is distributed throughout a business network. Thus the threat of a mediator sharing the transaction information with a third party is eliminated.  
\item {\em Use of Asymmetric Cryptography}.  The blockchain technology uses asymmetric key cryptography to protect the identity of transaction owners and to authenticate a transaction.   Transactions are delinked from the real-world identity of transaction owners. The transaction owners are identified through using public keys which an owner can generate multiple times.  In this way, transactions are pseudo-anonymous. The private keys are used to authenticate a transaction. 
\item  {\em Additional Mechanism for Anonymity}:  In the majority of blockchains implementations, transaction owners are identified through pseudo-anonymous identity. To provide full anonymity, additional mechanisms such as mixing transaction information, and a cryptographic technique called Zero-Knowledge proof can be used.  Zero-knowledge proof \cite{wang2018designing} is a verification technique which assures the validity of information without disclosing additional information.
\end{enumerate}

 \begin{sidewaystable}[h]
     \centering
     \begin{tabular}{|p{2.25in}| p{0.2in}| p{0.2in}|p{0.2in}|p{0.2in}|p{4.5in}|}
     \hline
     Functional Requirements   & \multicolumn{4}{|p{0.8in}|}{Blockchain Elements} & Remarks \\ \cline{2-5}
         & 
       \begin{sideways}DL\end{sideways}  & 
       \begin{sideways}CRYPT\end{sideways}  & 
       \begin{sideways}CP\end{sideways}  & 
       \begin{sideways} SC \end{sideways}  & 
      \\ \hline

        To enroll  students for examinations [FRQ1] & X & X & X &  &  This functionality shall record the cryptographic hash of records in the DL. The consensus protocol shall ensure consistency among all copies of DL.   \\ \hline 
       
        To assign identification credentials to students [FRQ2] & X  & X &  &  &  This functionality shall assign unique public keys to students using blockchain as a Public-Key-Infrastructure \cite{yakubov2018blockchain}.    \\ \hline 
       
       To issue hall tickets [FRQ4] & X & X &  X & & Hall tickets shall be implemented as tokens/coins which are transferred to the accounts of students before examinations and they are collected upon entry inside a hall.   \\ \hline 
     
     To generate  question papers for examinations [FRQ5] & X & & X & X & This functionality shall record hash of question item banks in the DL and implement the logic of automated generation of examinations as a smart-contract \cite{kale2013algorithm}. \\ \hline
     
    To distribute examinations related material [FRQ6]  & X & X & X & & Each stationery item shall be tagged with RFID tags so that the movement of stationery items can be tracked and traced. \\ \hline 
    
     To maintain examination record [FRQ7] & X & X & X & & Distributed ledgers shall be used to record the hash-values of the academic credentials of students. The consensus protocol shall ensure consistency among the multiple copies of ledgers. \\ \hline 
    
    To issue degree certificates and transcripts [FRQ8] & X & & & X & The functionality shall be implemented as a contract which generates certificates and transcripts for qualified students. \\ \hline 
     
    To verify and validate students' transcripts  and certificates [FRQ9] X &  &  &  & X & It shall be implemented as a smart contract. \\ \hline 
     
     \end{tabular}
     \caption{Mapping of Functional Requirements to Blockchain Architectural Elements}
     \label{frmap}
 \end{sidewaystable}
 
 \begin{sidewaystable}[h]
     \centering
     \begin{tabular}{|p{2.25in}| p{0.2in}| p{0.2in}|p{0.2in}|p{0.2in}|p{4.5in}|}
     \hline
     Information Security Requirements   & \multicolumn{4}{|p{0.8in}|}{Blockchain Elements} & Remarks \\ \cline{2-5}
         & 
       \begin{sideways}DL\end{sideways}  & 
       \begin{sideways}CRYPT\end{sideways}  & 
       \begin{sideways}CP\end{sideways}  & 
       \begin{sideways} SC \end{sideways}  & 
      \\ \hline 
    
    Data Confidentiality [SRQ1] & & X & & & Cryptography functions shall be used  to secure storage of information and communication among users.  \\ \hline 
     
   Data Integrity [SRQ2] & X & & X &  & Data integrity shall be   ensured through immutable DL and consensus protocols. \\ \hline
      
     Authentication [SRQ3] & & X & & & Blockchains cryptographic mechanisms have been increasingly employed to authenticate devices (e.g., IoT and  wireless sensor networks) \cite{hammi2018bubbles}. The proposed system shall use blockchain-based mechanisms for authenticating users and applications \\ \hline 
     
     Authorization [SRQ4]  & & & & X & A logic to ensure proper authorization and violation of authorization shall be implemented as a smart-contract \\ \hline 
     
     Denial of Service Attacks [SRQ5] & X &  &  X & & Blockchain-based systems being decentralized are more resilient against DoS attacks, and the level of resiliency depends on the consensus protocol used. The consensus protocol, such as  Practical Byzantine Fault Tolerant protocols shall be used to provide an acceptable level of resiliency. \\ \hline 
      
     Man-in-the-middle [SRQ6] &  & & & X &  A logic to detect and avoid such attacks  shall be implemented as a smart-contract. \\ \hline 
     
     Privilege escalation [SRQ7]  &  & & & X &  A logic to detect and avoid such attacks  shall be implemented as a smart-contract. \\ \hline
     
     \end{tabular}
     \caption{A Mapping of Non-Functional Requirements to Blockchain Architectural Elements}
     \label{nfrmap}
 \end{sidewaystable}

\subsection{Consensus Protocols}
In decentralized systems, agreeing upon the global state of the transaction is a challenge. In a centralized system, this is not an issue because only one copy of transaction history is present at the central authority (e.g., Banks main Server machine). Blockchain being a decentralized system holds multiple replicas of transactions at several nodes. Agreeing upon the unique state of the transaction is an issue which is solved by executing a consensus process involving all the nodes in the system. This process is typically carried out in three stages. In the first phase, a node is elected/selected as a leader node to decide upon a single state. In the second stage, transactions are validated. In the third stage, transactions are committed.

A variety of consensus algorithms exists \cite{sankar2017survey} in a blockchain-based system. These are often compared based upon how scalable the algorithm is and several malicious nodes it tolerates. The Proof-of-Work (PoW) algorithm used in Bitcoin is one example of the consensus protocol. It selects the leader node responsible for deciding upon a global state by solving a cryptographic puzzle.
The Proof-of-Stake (PoS) is another consensus protocol in which a leader is selected with the highest stakes in the network. The Practical Byzantine Fault Tolerant (PBFT) is the third example of consensus protocol for which the goal is to reach consensus in the event of failure of some of the nodes (e.g., up to 30\% nodes in the network) or they fail to provide accurate information to reach consensus.

\subsection{Smart-Contracts}
Smart-contracts \cite{cong2019blockchain} are the most significant element in the blockchain-based system because it provides mechanism to configure the behaviour of such systems. Blockchain programmers can customize the working of blockchain systems by writing programs called {\em Smart-Contract}. The smart contracts are scripts which are executed when a specific event occurs in a system. For example, in the context of Bitcoin, a coin may be released when more than one signatures are validated, or when miners solve a cryptographic puzzle.

These scripts can be written in a native language provided by blockchain systems or general-purpose programmable language. For example, Bitcoin provides a simple and less expressive native language to write a smart contract while Ethereum provides a Turing complete native language called Solidity to write smart contracts. In Hyperledger, blockchain programmers can write a smart contract in a general-purpose language such as Java/Go.

 \begin{figure}[t]
\centering
\begin{subfigure}{.5\textwidth}
  \centering
  \includegraphics[scale=0.24]{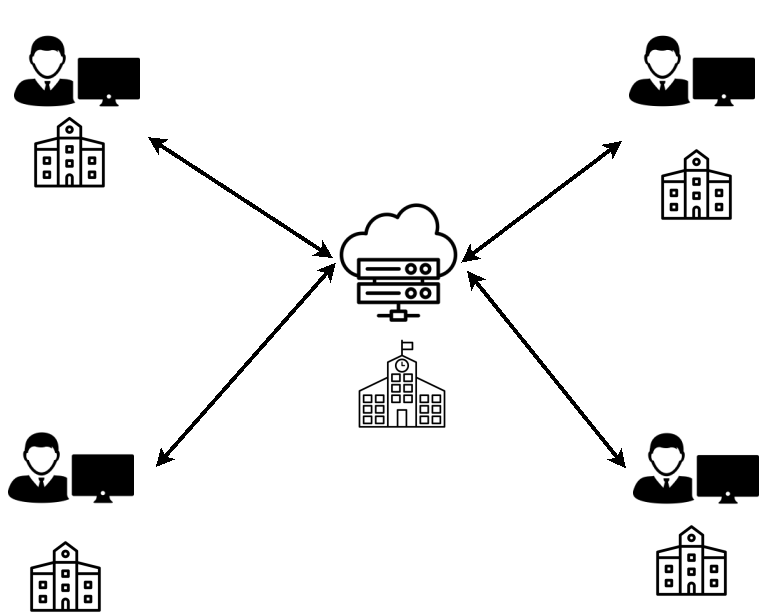}
  \caption{A Cloud-based Design}
  \label{fig:sub1}
\end{subfigure}%
\begin{subfigure}{.5\textwidth}
  \centering
  \includegraphics[scale=0.24]{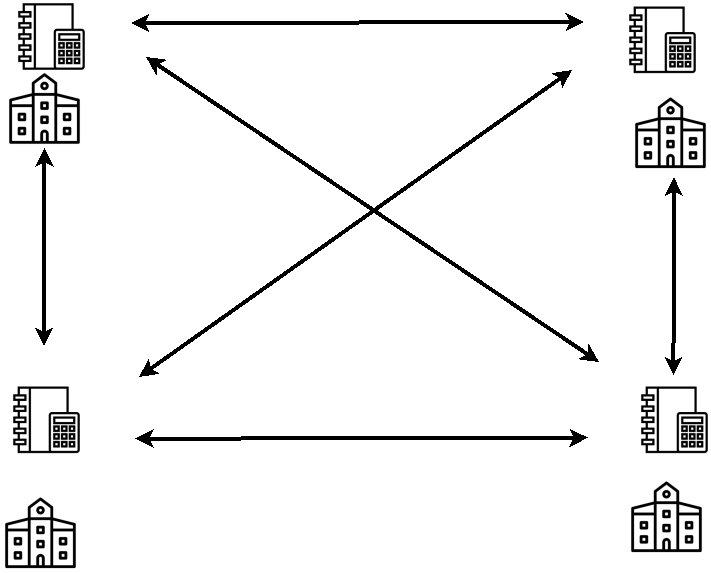}
  \caption{Decentralized and Autonomous Design}
  \label{fig:sub2}
\end{subfigure}
\caption{Centralized versus Decentralized University Examinations System}
\label{fig:test}
\end{figure}

 \section{A Prescriptive Architecture for University Examinations}
 
 An architecture of a software system can be designed by identifying Architecturally Significant Requirements (ASR) and mapping them to the elements in the desired architecture style \cite{kiwelekar2010ontological}. For example, a three-tier architecture of a system can be specified by mapping ASR to the {\em presentation}, {\em business process logic} and {\em data access layer}. Such kinds of architecture design are referred to as {\em prescriptive architecture}.  In addition to architectural mappings, the prescriptive architectures also include a set of essential design choices to be made early in the life-cycle. 
 The purpose of prescriptive architecture is to provide broad guidelines for downstream software engineering processes such as detailed design and implementation.  
 
\begin{figure}[t]
     \centering
     \includegraphics[scale=0.25]{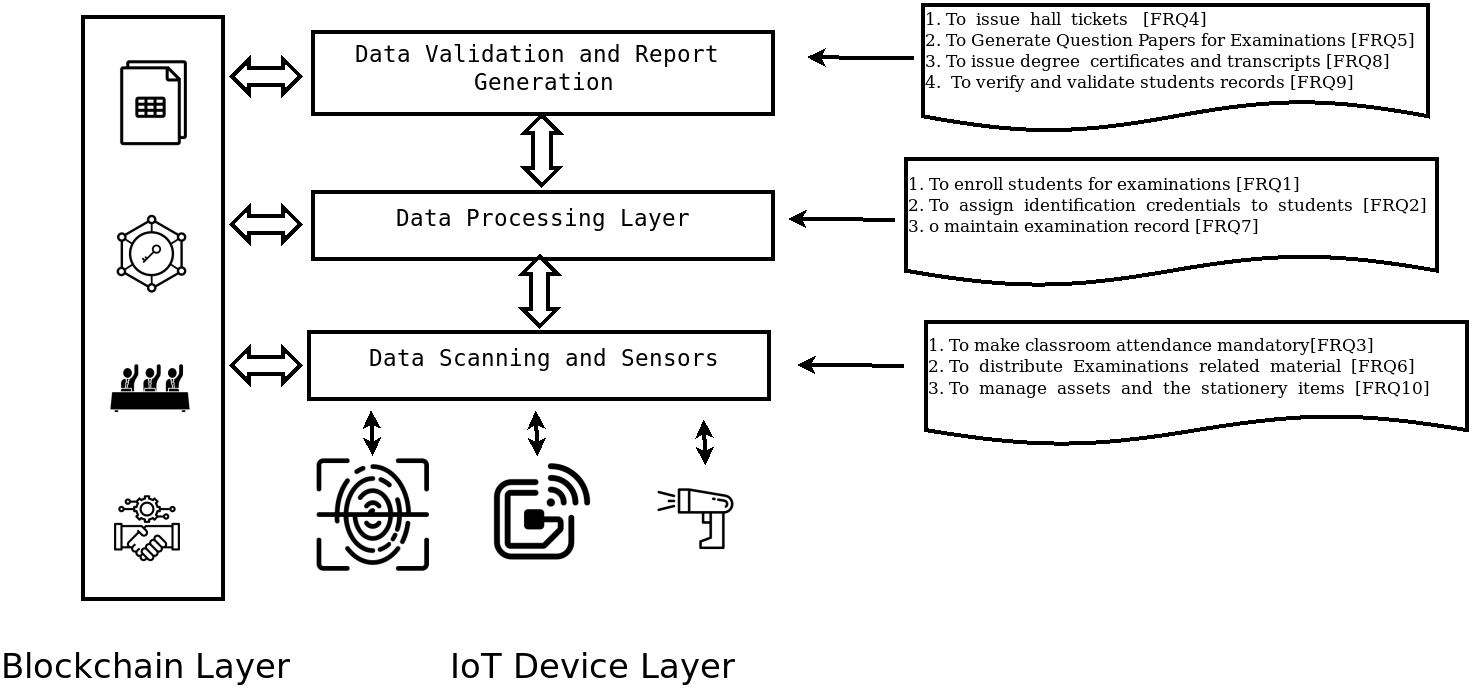}
     \caption{Software Architecture of the Proposed System for University Examinations}
     \label{layer}
 \end{figure}

 Figure \ref{layer} shows  the proposed software architecture designed around {\em Layered} architecture style. We prescribe a three-layered software organization with one additional cross-cutting layer. These layers are: {\em (i) Data Scanning and Data Sensing Layer}, {\em (ii)  Data Processing Layer}, {\em (iii) Data Validation and Report Generation Layer}. Most of the security requirements to be implemented using Blockchain Technology are {\em cross-cutting concerns} and represented as a vertical module. A cross-cutting concern is a requirement whose implementation is distributed
across multiple software modules \cite{eaddy2008crosscutting}.
 
 \subsection{IoT  for Monitoring and Asset Management} 
 
 We have grouped the functional requirements identified in Section \ref{fr} in two broad categories one that needs the IoT technology for the implementing requirements and others that need blockchain technology.  
 
 Our motivation behind using IoT is to support continuous monitoring and management of assets and inventory of stationery items required to conduct examinations addressing the requirements FRQ3 and FRQ10. 
 
 The following three IoT devices are used for this purpose.
 \begin{enumerate}
     \item {\bf Bio-metric Sensors for Attendance Monitoring} The proposed system shall collect the data about students' class attendance through bio-metric sensors such as finger-print scanners \cite{zainal2014design}.  The collected information shall be recorded updated on a daily/weekly basis to the blockchain ledgers. Based on the collected attendance information, the system shall issue admit cards to students satisfying minimum attendance requirements. Thus, it ensures the functional requirement FRQ3.
     \item {\bf Bar-code Scanner for data updating} Maintaining up-to-date information of stationery items required to conduct examination is one of the challenges. To address this challenge, the system shall use bar-code scanning and labelling of stationery items \cite{vamsi2020iot}. The inventory shall be maintained in a decentralized manner through blockchain technology.
     \item {\bf RFID for asset localization and management} The system shall use Radio Frequency Identifier (RFID) tags to locate and track the movement of mobile equipment and machinery such as printers, vehicles, laptops,  and server machines. The tracking information shall be recorded in a decentralized manner through the use of blockchain technology \cite{toyoda2017novel}.
 \end{enumerate}
 
 \subsection{Blockchain  Technology for  Decentralization and Autonomous Functioning}
 The choice of blockchain technology as the primary method for implementation is guided by the intention to bring more decentralization and autonomy in the functioning of university departments and colleges affiliated to it. This section provides a blueprint to implement such a system by defining a mapping of software requirements to the main blockchain elements.
 
 The Section \ref{fr} and \ref{nfr} identify such architecturally significant requirements which are functional and non-functional in nature. The major architectural elements in blockchain-based decentralized and autonomous systems are described in Section \ref{ae}. These are (i) Distributed Ledgers (DL), (ii) Cryptography (CRYPT), (iii) Consensus Protocols (CP), and (iv) Smart-Contracts (SC). 
 
 The first kind of mappings, as defined in Table \ref{frmap}, provides the mapping of functional requirements identified in Section \ref{fr} to the four elements of a blockchain-based system. The rationale behind a particular mapping and design guidelines are briefly described under the {\em Remark} column in the table. Similarly, Table \ref{nfrmap} provides mapping for the security requirements identified in Section \ref{nfr}.
 
 In addition to architectural mappings, a prescriptive architecture includes a discussion on design choices considered while specifying a high-level of design. Here we consider the following design choices.
 
 \begin{enumerate}
     \item {\bf Public vs Private Blockchain}: A business network implemented around public blockchains allows any member to join or leave the network without permission of the authority. For example, Bitcoin is a public blockchain, any node can participate during a consensus process.
     
     Unlike a public blockchain, membership of business network formed around the private blockchain is restricted, and it is granted by a membership service solely implemented for this purpose.

     We prefer a private blockchain over public blockchain because the university examinations system is small in size as compared to the community dealing with cryptocurrencies. Also, our purpose is to support existing administrative processes followed by universities. The role of universities is to grant membership to affiliated colleges which can be implemented as a membership service. Here, it also be noted that our purpose is not to create a self-governing digital university system around the blockchain platform.
     
   \item  {\bf Consensus Protocols} As seen in Section \ref{ae}, the purpose of consensus protocols is to reach consensus on the global state of ledgers distributed across the network. Various consensus protocols such as Proof-of-Work (PoW), Proof-of-Stake (PoS), and Practical Byzantine Fault Tolerant (PBFT) are some of the alternatives available for this purpose.
   
   We prefer PBFT over other consensus protocols because it provides fault-tolerant behaviour or it protects from a  single point failure situation. In addition to it, PBFT  is a scalable protocol tolerating about one-third malicious users. 
   
     \item {\bf Blockchain platform for implementation} Two existing blockchain frameworks, i.e. Hyperledger and Ethereum, are considered. The Hyperledger is preferred because of the available development support, and shorter learning time owing to the support of Java as a programming language for coding smart-contracts \cite{valenta2017comparison}. 
 \end{enumerate}

 \section{Conclusion}
 
 The chapter proposes the use of Blockchain technology and IoT to build administrative support for the university examinations system. We considered the examinations system as practised in Indian Universities. Many Indian universities have started employing cloud-based solutions to administer examinations conducted by them. However, cloud-based services are susceptible to attacks such as breaching the privacy of information shared to them and leakage of confidential information vital for fairly conducting examinations. Hence the motivation behind adopting a blockchain-based approach to administer examinations originates from the intention to enhance the reputation of universities by improving trust in the process of assessment.
 
To develop the high-level design of the Blockchain and IoT based approach, the chapter identifies architecturally significant requirements which include functional needs as well as information security requirements.  Then the identified requirements are mapped to the main architectural elements of the blockchain-based system, i.e. distributed ledgers, cryptography, consensus protocols and smart contracts. The use of  IoT devices such as Finger-print scanners, Bar code scanners, and RFID  are suggested for asset management and localization.

The approach proposed in the chapter has the potential to realize various administrative tasks during examinations in a hassle-free manner. Further, migrating existing software-based solutions used for administering university examinations to a blockchain-based solution is the biggest challenge that needs to be addressed at the time of implementation.
\bibliographystyle{plain}
\bibliography{ref}
\end{document}